\documentclass[lettersize,journal]{IEEEtran}
\usepackage[table]{xcolor}
\usepackage{amsmath,amsfonts}
\usepackage{algorithmic}
\usepackage{algorithm}
\usepackage{array}
\usepackage{textcomp}
\usepackage{stfloats}
\usepackage{url}
\usepackage{verbatim}
\usepackage{xspace}
\usepackage{graphicx}
\usepackage{cite}
\usepackage[T1]{fontenc}
\usepackage[utf8]{inputenc}
\usepackage{babel}
\usepackage[font=small,labelfont=bf]{caption}

\usepackage{soul}
\usepackage{ulem}

\usepackage{tikz}
\usepackage{multirow}
\usepackage{bookmark}
\usepackage{booktabs}
\usepackage{tabularx}
\usepackage{threeparttable}
\usepackage{arydshln} 
\usepackage{enumitem}
\usepackage{cleveref}

\usepackage{pifont}
\usepackage{caption}
\usepackage{subcaption}

\usepackage{changes}

\newcolumntype{\cX}[1]{>{\centering\arraybackslash}p{#1}}

\newcommand{\Sys}{Proteus\xspace}

\begin{document}

\title{\Sys: Simulating the Performance of \\ Distributed DNN Training}

\author{\rm{Jiangfei Duan}$^{\dag}$ \hspace{1.2em} \rm{Xiuhong Li}$^{\S\sharp}$ \hspace{1.2em} Ping Xu$^\ddag$ \hspace{1.2em} Xingcheng Zhang$^\sharp$\\
 \rm{Shengen Yan}$^\S$ \hspace{1.2em} Yun Liang$^{\S}$ \hspace{1.2em} Dahua Lin$^{\dag\sharp}$
\\ [.3em]
The Chinese University of Hong Kong$^\dag$ \hspace{1.2em} Shanghai AI Laboratory$^\sharp$ \\
Peking University$^\S$ \hspace{1.2em} SenseTime Research$^{\ddag}$
}

\maketitle

\begin{abstract}
DNN models are becoming increasingly larger to achieve unprecedented accuracy, and the accompanying increased computation and memory requirements necessitate the employment of massive clusters and elaborate parallelization strategies to accelerate DNN training. In order to better optimize the performance and analyze the cost, it is indispensable to model the training throughput of distributed DNN training. However, complex parallelization strategies and the resulting complex runtime behaviors make it challenging to construct an accurate performance model. In this paper, we present Proteus, the first standalone simulator to model the performance of complex parallelization strategies through simulation execution. Proteus first models complex parallelization strategies with a unified representation named \textit{Strategy Tree}. Then, it compiles the strategy tree into a distributed execution graph and simulates the complex runtime behaviors, \textit{comp-comm overlap} and \textit{bandwidth sharing}, with a \underline{H}ierarchical \underline{T}opo-\underline{A}ware \underline{E}xecutor (\textit{HTAE}).
We finally evaluate Proteus across a wide variety of DNNs on three hardware configurations. 
Experimental results show that Proteus achieves $3.0\%$ average prediction error and preserves order for training throughput of various parallelization strategies. Compared to state-of-the-art approaches, Proteus reduces prediction error by up to $133.8\%$.
\end{abstract}

\begin{IEEEkeywords}
Deep neural networks (DNNs), distributed training, parallelism, performance modeling, simulation.
\end{IEEEkeywords}

\section{Introduction}
\label{intro}

Recent years, progressively larger DNN models continue to break predictive accuracy records~\cite{he2016deep,krizhevsky2012imagenet,vaswani2017attention,radford2018improving,radford2019language,brown2020language,lepikhin2020gshard}.
As these models grow, they are becoming computationally and memory expensive to train. To efficiently train DNN models, large GPU clusters and sophisticated parallelization strategies are employed to accelerate the training process~\cite{dean2012large,mattson2019mlperf,shoeybi2019megatron,narayanan2021efficient,li2021terapipe,rajbhandari2020zero,ren2021zero,rajbhandari2021zero,fang2021patrickstar}.
For example, NVIDIA trained an 8.3 billion parameters language model on 512 GPUs with expert-designed hybrid data and model parallelism~\cite{shoeybi2019megatron}.

Since the training performance (throughput) of a DNN highly depends on its parallelization strategy, a natural question to ask is that
\textbf{can we accurately model the performance of any parallelization strategy for a specified cluster}. 
Modeling the performance of a parallelization strategy is crucial for performance optimization and
analysis.
1) Knowing the performance of a parallelization strategy can guide our optimization. Performance model can
be leveraged to locate the bottleneck of a parallelization strategy in manual optimization
and compare different parallelization strategies in automated parallelization systems~\cite{jia2019beyond, wang2019supporting, fan2021dapple, narayanan2019pipedream}.
2) Because implementing a parallelization strategy on current deep learning frameworks \cite{paszke2019pytorch, abadi2016tensorflow} is error-prone, labor-intensive and resource-costing, an accurate performance model can save lots of effort and resources in evaluating it.
3) Predicting the performance of a parallelization strategy in advance can help us analyze cloud service budgets without requiring GPU resources, such as how many machine hours or nodes to buy, thereby saving computing resources.

Plenty of performance modeling approaches have been proposed to predict the performance of DNN models, but none of them scale beyond hybrid data and model parallelism. 
Most of recent efforts to model the performance of DNN models are constrained to the scenario of single GPU. For example, various analytical models \cite{kothapalli2009performance, zhang2011quantitative, liu2021seer} that build with hardware metrics and learning-based models \cite{chen2018tvm, baghdadi2021deep} that learn from runtime statistics are presented to study the performance of GPU kernels.
In multi-GPU scenario, prior works \cite{pei2019iteration, yan2015performance, qi2016paleo} build analytical or profiling-based performance models for different DNN layers and predict training performance by summing up the computation and communication time of each layer. These approaches focus on a small subset of parallelization strategies and are not applicable to emerging parallelization strategies. 

Some automated parallelization approaches \cite{fan2021dapple, jia2019beyond, elango2021pase} also build performance models for distributed DNN training. For example, FlexFlow \cite{jia2019beyond} customizes a simulator to evaluate parallelization strategies in SOAP space. However, these works aim at searching optimal parallelization strategy for a DNN model instead of accurate performance modeling for general parallelization strategies. The usability and scalability are greatly limited due to their non-programmability of parallelization strategy and small strategy space.

We find two main challenges that hinder us from constructing accurate performance models for distributed DNN training. One challenge is \textbf{how to model \textit{complex parallelization strategies}}. 
Since different parallelization strategies own distinct computation and memory consumption characteristics, handcrafted strategies that are composed of various parallelization strategies at diferent levels are designed to accelerate DNN training, especially large DNN models \cite{shoeybi2019megatron,narayanan2021efficient,li2021terapipe,rajbhandari2020zero,ren2021zero,rajbhandari2021zero,fang2021patrickstar}. For example, Megatron-LM combines recomputation~\cite{chen2016training} and hybrid pipeline, data and model parallelism to train large transformer models~\cite{shoeybi2019megatron}.
The other challenge is \textbf{how to model \textit{complex runtime behaviors}}.
The underlying assumption of prior works \cite{yan2015performance, qi2016paleo, elango2021pase, jia2019beyond} is that \textit{the cost of a single operator only depends on its input and output tensor shape} and it does not hold when meeting complex parallelization strategies. During runtime, communication operators can be overlapped with computation operators to hide the cost of gradient synchronization, and communication operators in different communication groups may share bandwidth resources. Such optimization or sharing is not free but will increase the cost of these operators and thus decreasing the throughput of the entire DNN model. Therefore, an accurate performance model should be able to explicitly capture the optimizations of complex parallelization strategies and the overhead incurred due to complex runtime behaviors.

\begin{table}[t]
    \centering
    \small
    \setlength{\tabcolsep}{3pt}
    \caption{Comparison of \Sys and existing approaches. Approaches in \textbf{\textit{italics}} are automated parallelization frameworks, the others are performance modeling frameworks.}
    \label{tab:comparison}
    \begin{tabular}{p{1.85cm}| \cX{1.1cm} \cX{0.85cm} | \cX{1cm} \cX{1.1cm} | \cX{1.25cm}}
\hline

\multirow{3}{*}{Approach} & \multicolumn{4}{\cX{4.7cm}|}{\cellcolor{lightgray}\textit{Complex} Parallelization Strategy} & \cellcolor{lightgray} \textit{Complex} \\
& \multicolumn{2}{\cX{2cm}}{Operator-Level} &  \multicolumn{2}{\cX{2.1cm}|}{Subgraph-Level} & \cellcolor{lightgray} Runtime \\
\cline{2-5}
                          & Comp. & Mem. & Pipeline & Recomp. & \cellcolor{lightgray} Behavior \\
\hline
\textbf{\textit{DAPPLE}}~\cite{fan2021dapple} & Data & & \checkmark &  & \\
\textbf{\textit{FlexFlow}}~\cite{jia2019beyond} & SOAP & & & & \\
\hline
Yan \textit{et al.}~\cite{yan2015performance} & Hybrid & & & \\
Pei \textit{et al.}~\cite{pei2019iteration} & Data & & & \\
Paleo~\cite{qi2016paleo} & Hybrid & & & & \\
\Sys (ours) & \textbf{Shard} & \checkmark & \checkmark & \checkmark & \checkmark \\
\hline
    \end{tabular}
\end{table}

To address these challenges, we present \Sys, a standalone simulation framework that aims at accurately modeling the training throughput for distributed DNN training. \Cref{tab:comparison} highlights the advantages of \Sys against existing approaches.

First, we introduce a hierarchical tree structure, \textit{Strategy Tree}, to model complex parallelization strategies. 
We find parallelization strategies can be classified into operator- and subgraph-level strategies as a DNN graph is often divided into disjoint subgraphs, each of which is assigned to a group of devices.
The strategies at operator-level specify how the operators and tensors are split and mapped to devices, while the strategies at subgraph-level indicate how to schedule subgraphs (details refer to \Cref{sec:bg}). 
The hierarchical structure of strategy tree provides a unified representation for parallelization strategies at different levels and enables \Sys to model the huge and complex strategy space.

Second, we propose \textit{HTAE (Hierarchical Topo-Aware Executor)} to simulate complex runtime behaviors, which are ignored in prior works. We observe that runtime behaviors that have a significant impact on performance can be categorized into two types, \textit{comp-comm overlap} and \textit{bandwidth sharing}. HTAE simulates the schedule of subgraphs and operators to detect runtime behaviors during execution and adapts operator cost according to detailed cluster topology, thus capturing complex runtime behaviors of different operators.

Given a DNN model and \textit{Strategy Tree}, \Sys automatically compiles them into a distributed execution graph by splitting operators and tensors and inserting inferred collective communication operators. Afterwards, \Sys predicts the training throughput and OOM (Out-Of-Memory) error by mimicking the schedule and execution of the execution graph considering the cluster topology. 

In summary, we propose \Sys, to the best of our knowledge, the first standalone simulator to enable simulating complex parallelization strategies through fine-grained scheduling and simulation execution. We make the following contributions in building \Sys:

\begin{enumerate}[nosep, before=\leavevmode\vspace*{0\baselineskip}, leftmargin=*]
    \item We classify parallelization strategies into operator- and subgraph-level and formulate a unified parallelization space with \textit{Strategy Tree} to model complex parallelization strategies.
    \item We identify two types of runtime behaviors that affect performance: \textit{comp-comm overlap} and \textit{bandwidth sharing}, and introduce \textit{Hierarchical Topo-Aware Executor} to dynamically detect and model such behaviors.
    \item We evaluate Proteus across a wide variety of DNNs on 3 hardware configurations. 
    Experiments show that Proteus achieves $3.0\%$ average prediction error and preserves order for training throughput of various parallelization strategies. Compared to state-of-the-art approaches, Proteus reduces prediction error by up to $133.8\%$.
\end{enumerate}

\section{Background: Distributed DNN training}
\label{sec:bg}

DNNs are commonly represented as computation graphs in modern DL frameworks \cite{paszke2019pytorch, abadi2016tensorflow}, with nodes as operators and edges as tensors.  
Parallelizing a DNN involves parallelizing elements in the computation graph which can be categorized into two levels of parallelization strategies.

\paragraph{\textbf{Operator-Level Strategy}}
Operators and tensors can be partitioned for parallel execution on multiple devices. This partitioning is operator-level, which can be further categorized into \textit{computation parallelization} and \textit{memory optimization} based on corresponding computation and memory aspects.

\begin{figure}[tp] 
\centering
\includegraphics[width=0.43\textwidth]{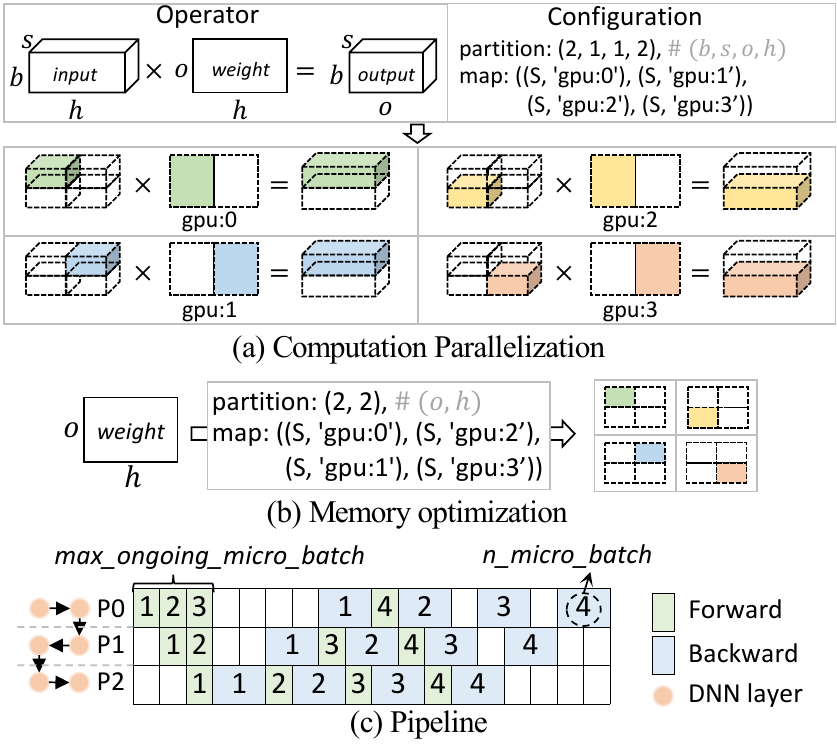}
\caption{Examples of parallelization strategies.}
\label{fig:psdemo}
\end{figure}

\textbf{Computation Parallelization} is achieved by partitioning the parallelizable dimensions of operators. Typically, we consider every unique dimension occurred in input or output tensors as parallelizable dimensions. We will describe different parallelization strategies taking the linear operator in Figure~\hyperref[fig:psdemo]{\ref{fig:psdemo}a} as an example:
$output(b, s, o) = \sum_{h}input(b, s, h) \times weight(o, h)$.
There are 4 unique dimensions: $b (batch)$, $s (sequence)$, $o (output\_channel)$, $h (hidden/reduction)$.

\textit{Data parallelism} is the most widely used parallelization strategy, which splits batch dimension ($b$) and replicates $weight$ on all devices.

\textit{Model parallelism} divides the operator in $o$ or $h$ dimension thus partitioning $weight$ into different parts and each part is trained on a dedicated device.

\textit{Hybrid parallelism} combines both \textit{data} and \textit{model parallelism} to partition operators.

\textit{Op shard} is a general parallelization strategy that exploits the power of partitioning arbitrary dimensions of~($b, s, o, h$). 

Figure~\hyperref[fig:psdemo]{\ref{fig:psdemo}a} shows an example configuration to \textit{shard} an operator in $b$ and $h$ dimensions. The \textit{partition} describes how to parallelize different dimensions and \textit{map} specifies how to place each partition. The operator is split into 4 ($|$\textit{partition}$|$) parts, with each assigned to a GPU. As reduction dimension ($h$) is partitioned, the operator produces 4 \textit{partial} output tensors, which should be aggregated to produce the final output tensor.

In this paper, \Sys targets on modeling the performance of \textit{general \textit{op shard}} unlike prior works that focus on \textit{data} and \textit{model parallelism}~\cite{pei2019iteration, qi2016paleo, yan2015performance}. SOAP \cite{jia2019beyond} partitions operators in $b, s, o$ dimensions and is a sub-space of op shard.

\textbf{Memory Optimization.} All dimensions of a tensor are parallelizable. Partitioning a tensor is achieved by splitting along its dimensions similar to partitioning operators. 

\textit{ZeRO} \cite{rajbhandari2020zero} and \textit{Activation partitioning} \cite{rasley2020deepspeed} partitions tensors in the first dimension and maps each part to a device to reduce redundancy. They can be combined with parallelization in other dimensions. Figure~\hyperref[fig:psdemo]{\ref{fig:psdemo}b} shows an example that partitions $o$ (ZeRO) and $h$ dimensions.

\Sys explicitly defines a parallelization strategy for each tensor in a DNN model. Figure~\hyperref[fig:psdemo]{\ref{fig:psdemo}a} shows that splitting an operator also creates implicit parallelization strategy for its input and output tensors. The inconsistency between the implicit and explicit strategy will incur additional communication (e.g. \textit{weight} need to transform from strategy of Figure~\hyperref[fig:psdemo]{\ref{fig:psdemo}b} to the implicit strategy of Figure~\hyperref[fig:psdemo]{\ref{fig:psdemo}a}).

\paragraph{\textbf{Subgraph-Level Strategy}}
A subgraph is composed of operators and tensors with dependencies. Parallelization strategies that describe the schedule of subgraphs are called subgraph-level strategies, including \textit{pipeline parallelism} and \textit{recomputation}, which balance training throughput and memory footprint by parallelizing subgraph computations. 

\textit{Pipeline parallelism} divides a computation graph into disjoint parts and assigns each part to a device group. It splits a batch input data into multiple micro-batches to exploit parallelism \cite{huang2019gpipe}. 
Figure~\hyperref[fig:psdemo]{\ref{fig:psdemo}c} shows a pipeline example with \textit{n\_micro\_batch} micro-batches for each subgraph. To reduce memory consumption, forward and backward micro-batches are interleaved~\cite{narayanan2019pipedream}, and \textit{max\_ongoing\_micro\_batch} limits the number of forward micro-batches on the flight.

\textit{Recomputation} (Activation Checkpointing) \cite{chen2016training} is a schedule that trades computation for memory. It frees forward subgraph activations after execution and recomputes when intermediate activations are required in backward pass.

Parallelization strategies at operator- and subgraph-level can be incorporated together thus formulating a complex parallelization space. \Sys leverages the hierarchical property to model complex parallelization strategies.

\section{\Sys Overview}

\begin{figure}[tp] 
\centering
\includegraphics[width=0.47\textwidth]{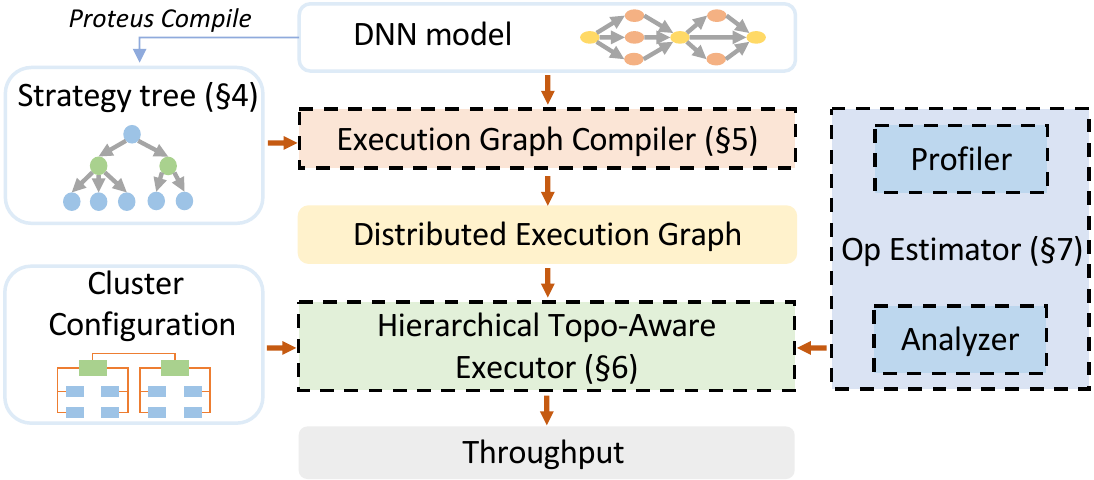}
\caption{An Overview of \Sys.}
\label{fig:overview}
\end{figure}

\begin{figure*}[tp] \centering{
\includegraphics[width=0.8\textwidth]{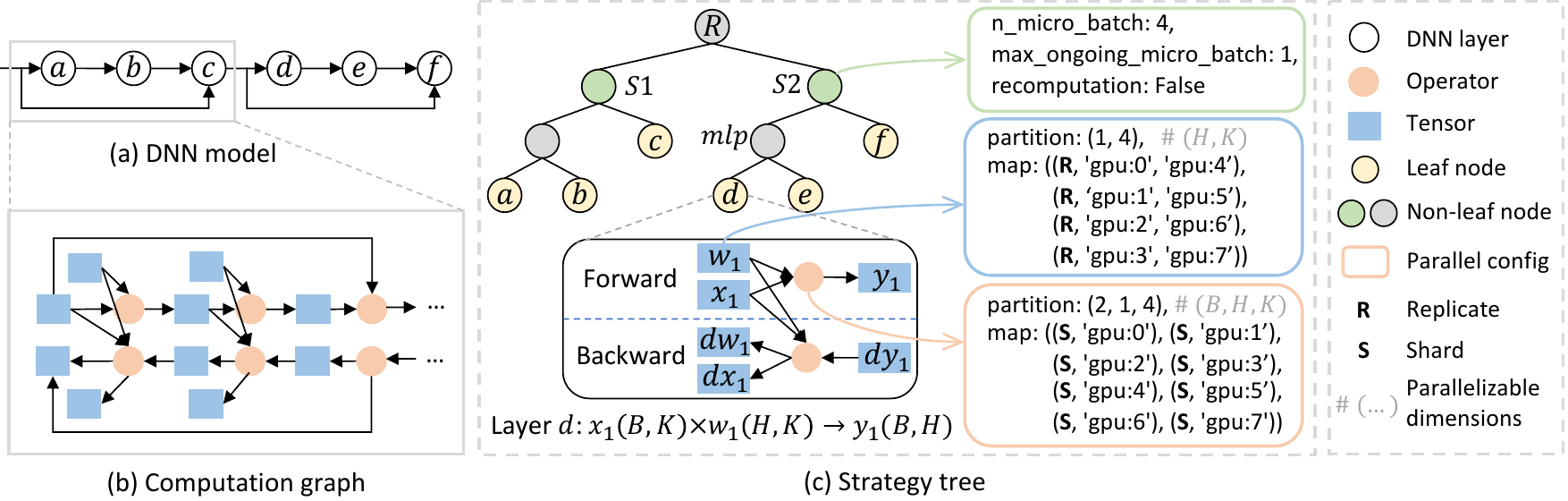}}
\caption{(a) A DNN model with 6 layers. (b) The computation graph of (a), including forward and backward computations. (c) The strategy tree of (a). Parallel configurations are assigned to non-leaf nodes and leaf nodes.}
\label{fig:stree}
\end{figure*}

Figure~\ref{fig:overview} shows an overview of \Sys, a simulation framework towards accurate performance modeling for distributed DNN training. Since the performance of distributed DNN training highly depends on the parallelization strategy, explicitly modeling a parallelization strategy is the first step for a performance model. \Sys uses a unified representation \textit{strategy tree} to model complex parallelization strategies (\Cref{sec:stree}). 

\Sys's \textit{execution graph compiler} (\Cref{sec:compiler}) bridges the gap between high level parallelization strategy and low level execution. It takes DNN model and strategy tree as inputs, and compiles DNN layers into tensors and operators. The compiler automatically inserts communication operators between tensors and generates a distributed execution graph.

In \Cref{sec:executor}, we first discuss the characterization and impact of runtime behaviors, then introduce \Sys's \textit{hierarchical topo-aware executor}, which simulates the schedule of the execution graph and predicts the training throughput. During simulation, it adapts operator cost, which is first obtained with the \textit{op estimator} (\Cref{sec:impl}), considering the cluster configuration and dynamic runtime behaviors. 

\section{Strategy Tree}
\label{sec:stree}

This section introduces \textit{strategy tree}, a unified representation to model complex parallelization strategies, including both operator- and subgraph-level strategies. 

\Cref{fig:stree} shows a DNN model and its computation graph and strategy tree. Computation graph is commonly used to represent data dependencies between operators and tensors. However, the non-hierarchical structure makes it hard to distinguish different subgraphs, thus failing to model subgraph-level strategies (e.g. the subgraph-level strategy for $a$ and $b$ can only be assigned when we manually split the graph or assign strategies for all tensors and operators).

To solve the problem, \Sys models parallelization strategies with a hierarchical tree structure. Tensor, operator and subgraph are basic elements of a parallelization strategy, regardless of data dependencies. The tree structure provides a good abstraction for modeling strategies at different levels and capturing nested structure between various elements. \Sys models tensors and operators in leaf nodes and subgraphs in non-leaf nodes. The leaf and non-leaf nodes make it easier to determine operator- and subgraph-level strategies. A complete parallelization strategy consists of \textit{parallel configurations} on all tree nodes. We will further discuss and compare strategy tree with prior works in \S\ref{subsec:stree_cmp}.

\subsection{Tree Representation}
A leaf node models the forward and backward computation graphs of a DNN layer. As illustrated in Figure~\hyperref[fig:stree]{\ref{fig:stree}c}, leaf node captures all the forward and backward operators and the tensors they produce and consume. \Sys models tensors by their shape, and operators by a set of unique parallelizable dimensions extracted from input and output tensors (\Cref{sec:bg}).

A non-leaf node models a subgraph, which represents the forward and backward computation graphs of several DNN layers. It is possible to group different layers to create various subgraphs, which is a natural hierarchical structure. For example in Figure~\hyperref[fig:stree]{\ref{fig:stree}c}, the layers $d, e$ and $d, e, f$ constitute two non-leaf nodes respectively at different levels, and the root node models the whole DNN model.

\subsection{Parallel Configuration}
\label{subsec:pconfig}
The parallel configuration defines how different components are parallelized.
Operator-level strategies are specified with \textit{computation} and \textit{memory config} in leaf nodes, subgraph-level strategies are assigned to non-leaf nodes with \textit{schedule config}. 

\textbf{Computation/Memory Config.}
Computation (memory) configs are assigned to operators (tensors) in leaf nodes. It contains two aspects: \textit{partition} and \textit{map}. The \textit{partition} ($\mathcal{P}$) defines the degree of parallelism in each dimension and splits the operator (tensor) into $|\mathcal{P}|$ disjoint parts. Each part will be mapped to one or more devices defined by \textit{map}, namely shards on one device or replicates on a device group. In Figure~\hyperref[fig:stree]{\ref{fig:stree}c}, the computation config partitions the $B$ and $K$ dimensions of the forward operator into 2 and 4 parts respectively, and shard each part on one GPU device.

Memory config defines the real placement of a tensor. With this separated memory config, \Sys is able to express the space of memory optimization.

\textbf{Schedule Config.}
Schedule config specifies the subgraph-level strategy of a subgraph, with only one config needed for each non-leaf node due to the dual structure of the forward and backward subgraphs. The config has three aspects (Figure~\hyperref[fig:stree]{\ref{fig:stree}c}): \textit{n\_micro\_batch} denotes the number of micro-batches consumed by the subgraph, Since executing forward micro-batches increases memory consumption, \textit{max\_ongoing\_micro\_batch} limits the maximum number of forward micro-batches executed before each corresponding backward micro-batch at any time, and \textit{recomputation} indicates whether to use activation checkpointing.

Non-leaf nodes on the tree have a schedule config that is propagated from the parent node unless explicitly defined by the user. In particular, the schedule config on a non-leaf node is independent of the configs on leaf nodes. Strategy propagation will be discussed in \Cref{sec:impl}.

\begin{figure*}[t] \centering{
\includegraphics[width=0.88\textwidth]{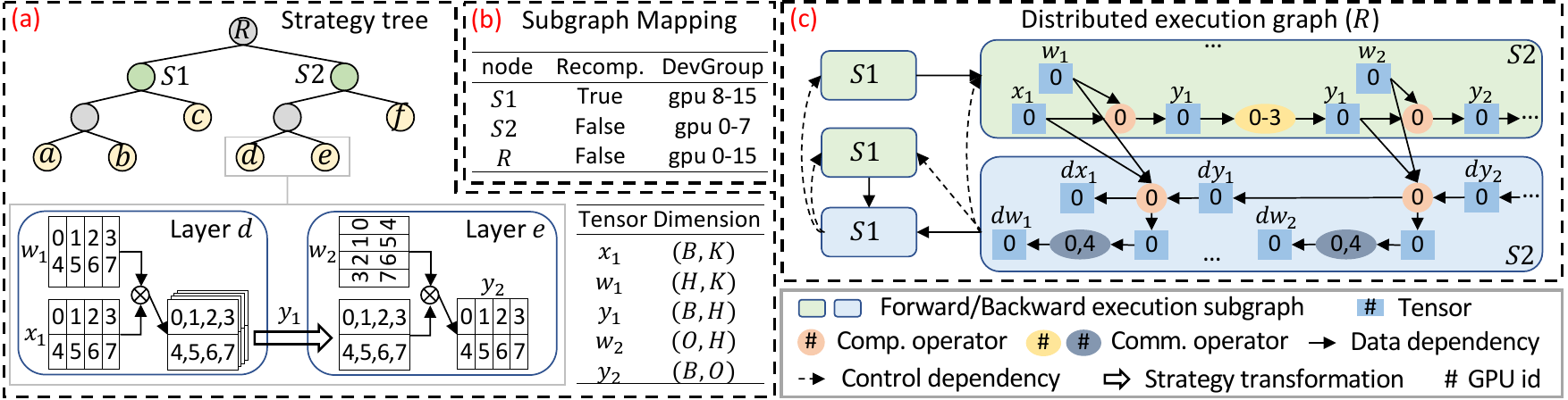}}
\caption{The strategy tree and corresponding distributed execution graph of DNN model in \Cref{fig:stree}.}
\label{fig:compiler}
\end{figure*}

\subsection{Discussion and Comparison}
\label{subsec:stree_cmp}

Existing frameworks \cite{fan2021dapple, jia2019beyond, zheng2022alpa, unity} relies on computation graph to unify the parallelization space including computation, memory and schedule. However, \Sys focuses accurate performance modeling for a parallelization strategy, while these frameworks are designed for automated parallelization and they will search over all combinations of subgraphs. Explicitly modeling parallelization strategies is necessary for performance modeling, but it is quite hard to specify a complex parallelization strategy for a DNN model in prior automated parallelization works.

GSPMD~\cite{xu2021gspmd} develops powerful programming APIs to specify parallelization strategies for different DNN layers, but the subgraph-level strategy is only supported for identical DNN blocks with tailored \textit{vectorized\_map} API. Furthermore, changing parallelization strategy also takes great efforts to rewrite the model. Our strategy tree unifies parallelization strategies at different levels and decouples parallelization strategy from model expression. By adjusting the strategy tree instead of the DNN model, \Sys can change the parallelization strategy for a DNN.

\section{Execution Graph Compiler}
\label{sec:compiler}

This section describes \Sys's \textit{execution graph compiler}, which connects high level parallelization strategies with low level execution. Given a strategy tree, the compiler creates a distributed execution graph by splitting tensors and computation operators and inserting communication operators and control dependencies. 

\subsection{Graph Compilation}

\Cref{fig:compiler} illustrates the workflow of execution graph compiler. 
\Sys first divides the DNN model into disjoint subgraphs based on its \textit{DevGroup}, which defines a set of devices, in order to parallelize the computations of different micro-batches. The DevGroup of a tree node is composed of all of its children nodes' DevGroups. \Sys splits all divisible nodes in breadth-first order from root node and a node cannot be divided unless all of its children nodes share some devices. Figure~\hyperref[fig:compiler]{\ref{fig:compiler}b} shows the DevGroups of three nodes. The DevGroup of node $S2$ is "gpu 0-7"  because layer $d$ and $e$ are partitioned and mapped to these devices. The root node $R$ is divided into 2 subgraphs since node $S1$ and $S2$ share no devices and they are not divisible. 

\Sys then compiles each subgraph into a forward and backward execution subgraph as showed in Figure~\hyperref[fig:compiler]{\ref{fig:compiler}c}.
Tensors and operators are split into small partitions such that each partition resides on and is executed by one device.
Communication operators, data and control dependencies are added to ensure the computational equivalence.

\textbf{Data dependency.}
Each tensor and operator has a parallel configuration that defines the partition and mapping, as discussed in \Cref{subsec:pconfig}. Due to the data dependency between tensors and operators, \Sys can infer a parallel configuration for each input and output tensor of operators. Once the two parallel configurations of a tensor are inconsistent, \Sys automatically inserts communication operators via \textit{strategy transformation} to adjust the parallel configuration, otherwise \Sys reuses original tensor partitions. Figure~\hyperref[fig:compiler]{\ref{fig:compiler}} shows the strategy transformation of tensor $y_1$. Layer $e$ partitions $y_1$ into 2 parts and each part replicates on 4 GPUs, but layer $d$ partitions $y_1$ into 8 partial tensors on 1 GPU. \Sys adds communication operators between $y_1$ in the execution subgraph of $S$2 to handle this 
inconsistency. 

For a subgraph with recomputation enabled, \Sys compiles it into two forward and one backward execution subgraphs and adjust the data dependency accordingly. The backward subgraph depends on one forward subgraph (i.e. recomputation subgraph) and the other subgraph can be immediately released after execution~(e.g. $S$1 in Figure~\hyperref[fig:compiler]{\ref{fig:compiler}c}).

\textbf{Control dependency.}
Control dependencies are inserted between execution subgraphs to follow the training schedule defined by the schedule config in non-leaf nodes. First, the forward subgraphs are control dependent on their corresponding backward subgraphs to limit peak memory consumption. Second, \Sys also adds control dependency for recomputation subgraphs such that they are executed immediately before the backward subgraphs. In Figure~\hyperref[fig:compiler]{\ref{fig:compiler}c}, node $S$1 has two forward subgraphs and one of them is control dependent on the backward of node $S$2.

\begin{figure*}[t]
\centering
     \begin{subfigure}[b]{11.5cm}
         \centering
         \includegraphics[height=2.5cm]{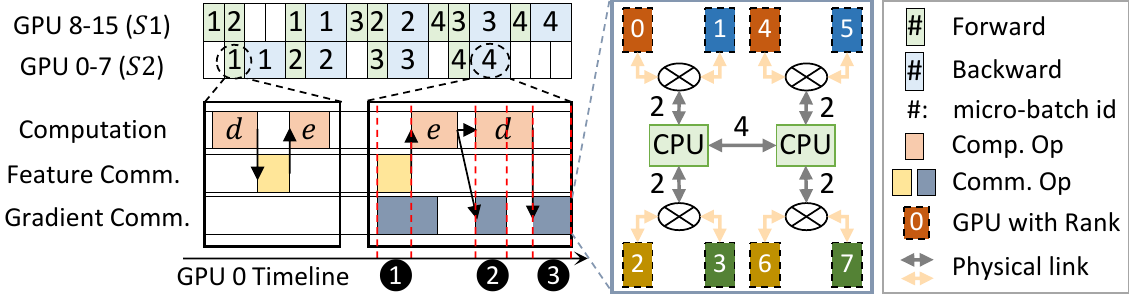}
         \caption{Execution timeline.}
         \label{fig:timeline_a}
     \end{subfigure}
     \quad
     \begin{subfigure}[b]{4.5cm}
         \centering
         \includegraphics[height=2.5cm]{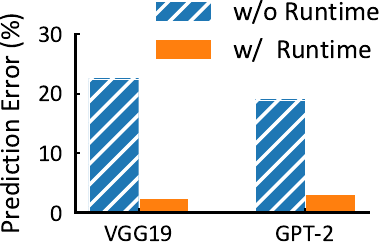}
         \caption{Prediction error.}
         \label{fig:timeline_err}
     \end{subfigure}
\caption{(a) The execution timeline and runtime behaviors of the distributed execution graph in \Cref{fig:compiler}. The number on gray link denotes the number of groups that share the link. (b) Comparison of prediction errors for modeling runtime behaviors or not.}
\label{fig:timeline}
\end{figure*}

\subsection{Strategy Transformation}
Strategy transformation converts tensors to desired parallel configurations with appropriate communication primitives. \Sys automatically infers collective communication primitives (e.g. All-Reduce \cite{nccl}), failing over to point-to-point communication if necessary. Since different primitives features distinct communication patterns, \Sys uses pattern matching to infer collective communication operators and corresponding communication groups. 

\Sys currently supports commonly used communication primitives in modern DL frameworks \cite{paszke2019pytorch, abadi2016tensorflow}. Prior work proposes new communication primitives, such as hierarchical reduce~\cite{2d-reduce, hierarchical-reduce} and CollectivePermute~\cite{xu2021gspmd}, to accelerate the communication. \Sys can be extended by including more candidate patterns.

\section{Hierarchical Topo-Aware Executor}
\label{sec:executor}

This section describes \Sys's \textit{Hierarchical Topo-Aware Executor} (HTAE), which simulates the schedule and runtime behaviors of a distributed execution graph and predicts the training throughput. 

\subsection{Performance Characterization}
Before introducing the design of HTAE, we first characterize the performance of distributed DNN training using the example of \Cref{fig:timeline_a}, which illustrates the execution timeline and runtime behaviors of \Cref{fig:compiler}. The forward and backward execution subgraphs are interleaved and the execution of $S1$ and $S2$ is parallelized on different GPU groups. Operators in Figure~\hyperref[fig:compiler]{\ref{fig:compiler}c} consist of three types that can be executed simultaneously: computation, feature and gradient communication operators, and they are scheduled into three streams following data dependency. Modeling the training performance is to model the execution timeline, including schedule, computation and communication.

\textbf{Runtime Behavior.}
Prior work \cite{yan2015performance, pei2019iteration, qi2016paleo, jia2019beyond} assumes that the operator cost is fixed and focuses on modeling the performance of single operator. The training speed of a DNN is the summation of all the operators' costs. However, \textit{runtime behavior}, which is ignored in prior work, has emerged as a critical aspect determining training performance under today's sophisticated parallelization strategies and optimizations. It is crucial to model runtime behaviors towards an accurate performance predictor since they can affect the execution cost of operators. \Cref{fig:timeline_err} shows that ignoring runtime behaviors results in large prediction error on a cluster with 32 GPUs.

We find that major runtime behaviors can be categorized into two types. First, \textit{bandwidth sharing} describes the scenarios that different communication operators compete for bandwidth (\Cref{fig:timeline_a}\ding{182},\ding{184}). Second, \textit{comp-comm overlap} refers to the overlap of computation and communication operators (\Cref{fig:timeline_a}\ding{183}). In addition, different computation operators could be overlapped on single GPU~\cite{ios_hansong}, \Sys does not model such scenario since it is rarely used in distributed DNN training. \Cref{fig:timeline_a}\ding{184} shows an example of bandwidth sharing by mapping gradient communication operators to a single node machine. The gradient communication includes 4 groups indicated by the GPU color: \{\{0, 4\}, \{1, 5\}, \{2, 6\}, \{3, 7\}\}, and their costs rise due to the competition for available bandwidth of scarce physical links. 

\Sys is the first system to study and model runtime behaviors for distributed DNN training. Unlike prior analytical frameworks \cite{yan2015performance, qi2016paleo}, \Sys predicts training performance via simulation since runtime behaviors only occur during execution. 

\subsection{Simulator Design}

\begin{figure}[tp] 
\centering
\includegraphics[width=0.42\textwidth]{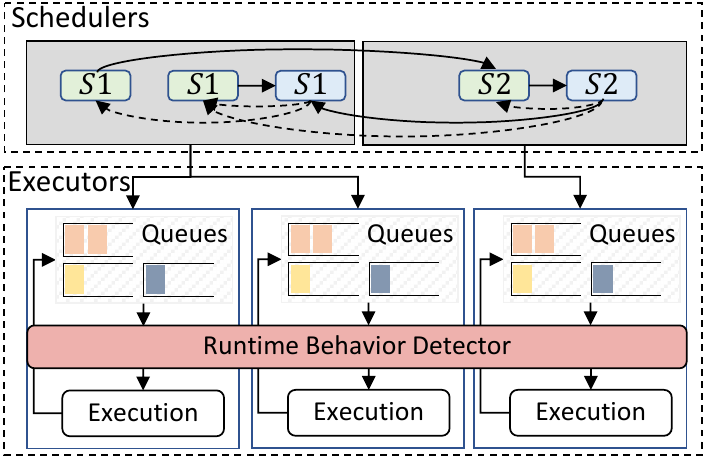}
\caption{The design of hierarchical topo-aware executor.}
\label{fig:exec}
\end{figure}

\Cref{fig:exec} shows the design of \Sys's two level simulator, HTAE. The first level is \textit{scheduler}, which consists of several second level \textit{executors}. Different \textit{schedulers} can time-share \textit{executors}. To predict the performance, HTAE first gets single operator cost with \textit{op estimator} (\Cref{sec:impl}) and then simulates the schedule of subgraphs and operators to discover runtime behaviors. During simulation, operator cost are adapted to model runtime behaviors considering the cluster topology.

\textbf{Cluster Configuration.}
Cluster Configuration describes the topology of training cluster. There are two types of configurable parameters in device topology. For intra-node topology, we can set device type, device memory, number of devices in a node and the intra-node connection, which describes the physical connections among devices (e.g. GPUs and CPUs). For inter-node topology, we can specify the number of nodes and inter-node connection bandwidth.

\textbf{Scheduler.}
Each scheduler is assigned several forward and backward execution subgraphs, and it interleaves the execution of them based on data and control dependencies to balance micro-batch parallelism and peak memory consumption. The scheduler first selects current execution state (forward or backward), then it chooses one subgraph from available dependency-free execution subgraphs. It alternates different backward subgraphs and prefers forward subgraph that enables backward execution. After determining the subgraph to be executed, the scheduler dispatches initial tasks to executors and begin executing.

\textbf{Executor.}
The \textit{executor} schedules the execution of operators for a subgraph and records the peak memory consumption. Each executor contains a computation queue, a feature communication queue and a gradient communication queue (\Cref{fig:exec}). Operators in different queues can be executed concurrently such that achieving \textit{comp-comm overlap}. By separating feature and gradient communication queue, \Sys makes it possible to overlap feature and gradient communication and avoid feature communication blocked by gradient communication.

The executor executes computation and communication alternatively. It pops a computation operator from the queue for computation execution and pops one feature and one gradient communication operator at the same time for communication execution. These operators are first sent to the \textit{runtime behavior detector} to check runtime behaviors and executed afterwards. During execution, the operator cost will be accumulated to count the time cost for each queue separately. The execution of operators will decrease the number of dependencies of their consumers and dependency-free operators will be put into the corresponding queue.

\textbf{Memory Consumption.}
\Sys predicts whether a parallelization strategy will out-of-memory (OOM) by monitoring the memory consumption of executors. During execution, each operator reads and writes some tensors. HTAE monitors executor memory footprint by recording these tensor activities. When writing a new tensor, HTAE tracks its memory consumption and reference counter. The memory will be released when the reference counter decreases to zero.

\subsection{Modeling Runtime Behaviors}
As previously discussed, the operator cost may change during execution due to complex runtime behaviors. The \textit{runtime behavior detector} checks runtime behaviors for all operators and adapt operator cost accordingly. To enable efficient detection, it keeps execution history records of different execution streams.

\textbf{Bandwidth Sharing.}
There are two types of bandwidth sharing. One is inside a group of gradient or feature communication operators (\Cref{fig:timeline_a}\ding{184}), and the other is between a group of gradient and feature communication operators (\Cref{fig:timeline_a}\ding{182}). These operators transfer datas within different device groups and compete for bandwidth of shared physical links. To model this behavior, \Sys assumes that concurrent operators fairly share the bandwidth of a physical link and detects how many communication groups share a link during execution. We find this assumption generally holds in practice.

\begin{figure}[tp] 
\centering
\includegraphics[width=0.27\textwidth]{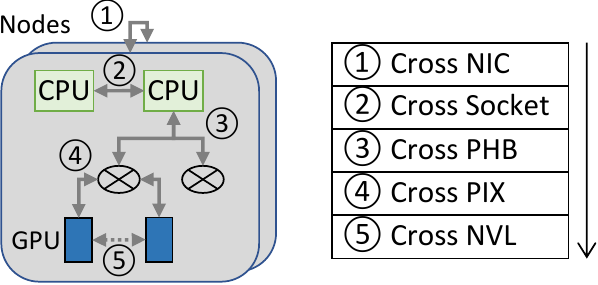}
\caption{The hierarchy of bandwidth sharing.}
\label{fig:bw_share}
\end{figure}

\Sys first checks bandwidth sharing for feature and gradient communication operators separately by mapping communication groups to cluster topology. \Cref{fig:bw_share} shows the hierarchy of physical links in a cluster and \Sys detects bandwidth sharing following this hierarchy. \Sys starts from NIC bandwidth sharing. Each communication group is split into sub-groups such that each sub-group is composed of devices in the same node. The groups that consist of more than two sub-groups fairly share the bandwidth of NIC. \Sys checks all the physical links from top to bottom.

\Sys finally detects the intersection effects of feature and gradient communication groups. Since the communication volumes and operations of these groups may different, \Sys only adapt operator cost for the overlapped parts. The detection algorithm is the same as the first step, except for that the communication groups include both feature and gradient communications.

\textbf{Comp-Comm Overlap.}
In distributed DNN training, computation and gradient communication operators may overlap, because gradient communication operators can be launched asynchronously and feature communication operators usually block the computation stream. To detect \textit{comp-comm overlap}, \Sys keeps the start and end time of operators. When executing a computation (communication) operator, \Sys considers it \textit{overlapped} if it finds a gradient communication (computation) operator in execution. 

\Sys introduces an overlap factor $\gamma$ to model the effect of \textit{comp-comm overlap}. When finding an operator \textit{overlapped}, its cost will increase by $\gamma$. This is motivated by the observation that operator costs increase by about the same percentage on average during overlap.

To obtain $\gamma$, we profile the speeds of backward pass with and without overlapping in data parallel training and $\gamma$ is set to the increase ratio. As $\gamma$ is fixed for the type of machine and DNN model, we can get $\gamma$ in advance with few cost. Prior works~\cite{qiao2021pollux, yang2021perfestimator} also try to model the effect of \textit{comp-comm overlap}. For example, Pollux~\cite{qiao2021pollux} introduces a learnable parameter $\eta$ to model the data parallel training speed by combining computation time $T_{grad}$ and gradient communication time $T_{sync}$ with $({T_{grad}}^\eta + {T_{sync}}^\eta)^{1/\eta}$. These works target on data parallel training, and we find our simple formulation works pretty well with complex parallelization strategies (discussed latter in \Cref{ablation}).

\section{Implementation}
\label{sec:impl}

\Sys is implemented as a standard python library ($\sim9$K LoC). \Sys follows PyTorch \cite{paszke2019pytorch} API to build model.

\textbf{Construction of Strategy Tree.}
DNN models consists of modules that perform operations on data. A module is roughly equivalent to a DNN layer, and different modules can be nested together to construct a complex module. \Sys exploits the module structure to create a strategy tree from top to bottom in depth-first manner. Non-leaf nodes correspond to complex modules and leaf nodes correspond to layers. The root node represents the entire DNN model. \Sys tracks the construction of modules/layers and creates corresponding non-leaf/leaf nodes on the tree. The construction of strategy tree preserves the structure of DNN model and makes it easier to specify parallelization strategy for nodes.

\textbf{Strategy Propagation.}
\label{subsec:cfg_prop}
\Sys develops a strategy propagation algorithm to ease the programming difficulty of parallelization strategies. For a complete parallelization strategy, programmers are required to specify parallel configurations for critical leaf and non-leaf nodes. \Sys will propagate the parallel configurations to the other nodes.

\Sys first propagates parallel configurations from top to bottom following tree structure. The schedule config of a non-leaf node is inherited from its parent node unless explicitly defined. \Sys then propagates parallel configurations among leaf nodes following data dependency. The propagation proceeds in topological order and includes two steps: forward graph propagation and backward graph propagation. \Sys infers the memory config of a tensor according to its producer's computation config and infers the computation config of an operator according to its inputs' memory config. 

\begin{table}[t]
\centering
\small
\caption{Overview of the six benchmark models evaluated.}
\label{table:bench}
\begin{tabular}{ cccc }
\hline
Task & Model & \#Params & Dataset \\
\hline
\multirow{3}{*}{Vision} & ResNet50  \cite{he2016deep}     & 25.6M & \multirow{6}{*}{Synthetic} \\
                        & Inception\_V3\cite{szegedy2016rethinking} & 23.8M &  \\
                        & VGG19 \cite{simonyan2014very}        & 137M  &  \\
\cline{1-3}
\multirow{2}{*}{NLP}    & GPT-2  \cite{radford2019language}       & 117M &  \\
                        & GPT-1.5B \cite{radford2019language}      & 1.5B &  \\
\cline{1-3}
    Recommendation      & DLRM \cite{naumov2019deep}          & 516M &  \\
\hline
\end{tabular}
\end{table}

\textbf{Op Estimator.}
\label{subsec:op_est}
The \textit{op estimator} predicts the operator cost for all operators in the distributed execution graph. It contains a profiler and analyzer. Since \Sys focuses on modeling runtime behaviors, the profiler obtains the time cost of computation operators by profiling them on target hardware, which costs little. There are lots of performance models to estimate single operator cost \cite{kothapalli2009performance, zhang2011quantitative, liu2021seer}. \Sys can be extended to adopt such models.

\begin{table}[t]
\centering
\small
\caption{Overview of hardware configurations evaluated.}
\label{table:hw_cfg}
\begin{tabular}{ ccccc }
\hline
Config & \#Node & \#GPU per Node & Intra-node & Inter-node \\
\hline
\textit{HC1} & 1 & 8$\times$TitanXp & PCI-e &  N/A \\
\textit{HC2} & 4 & 8$\times$V100 & NVLink &  100 Gbps \\
\textit{HC3} & 2 & 8$\times$A100 & NVLink &  200 Gbps \\
\hline
\end{tabular}
\end{table}

The analyzer estimates communication cost with \textit{$\alpha$-$\beta$} model \cite{alpha-beta-model}. It estimates the bandwidth of a communication group according to the detailed cluster topology. When estimating the time cost of a collective operation, a correction factor is applied to revise the bandwidth to reflect the characteristics of different collective operations. 
To simplify implementation, we utilize NCCL topo detection algorithm \cite{nccl} to find all the communication channels of a communication group and its bandwidth is the summation of these channels. 

\section{Evaluation}

\subsection{Methodology}
All the experiments are conducted with PyTorch 1.8 (CUDA 10.1, cuDNN 7.6.5 and NCCL 2.7.8).

\noindent
\textbf{Benchmarks.} Table~\ref{table:bench} summarizes the six representative DNN models that we used as benchmarks, they are widely used in prior works \cite{qi2016paleo,jia2019beyond,liu2021seer,fan2021dapple}. We evaluate throughput with synthetic dataset, which ignores the data loading latency.
Modeling real-world dataset is orthogonal to \Sys.

\noindent
\textbf{Hardware Configurations.} \Sys is evaluated across three different hardware configurations. \Cref{table:hw_cfg} summarizes the cluster type and size, intra- and inter-node connections.

\subsection{Simulation Accuracy}
\label{sec_sim_acc}

\begin{figure*}
    \centering
    \includegraphics[width=\textwidth]{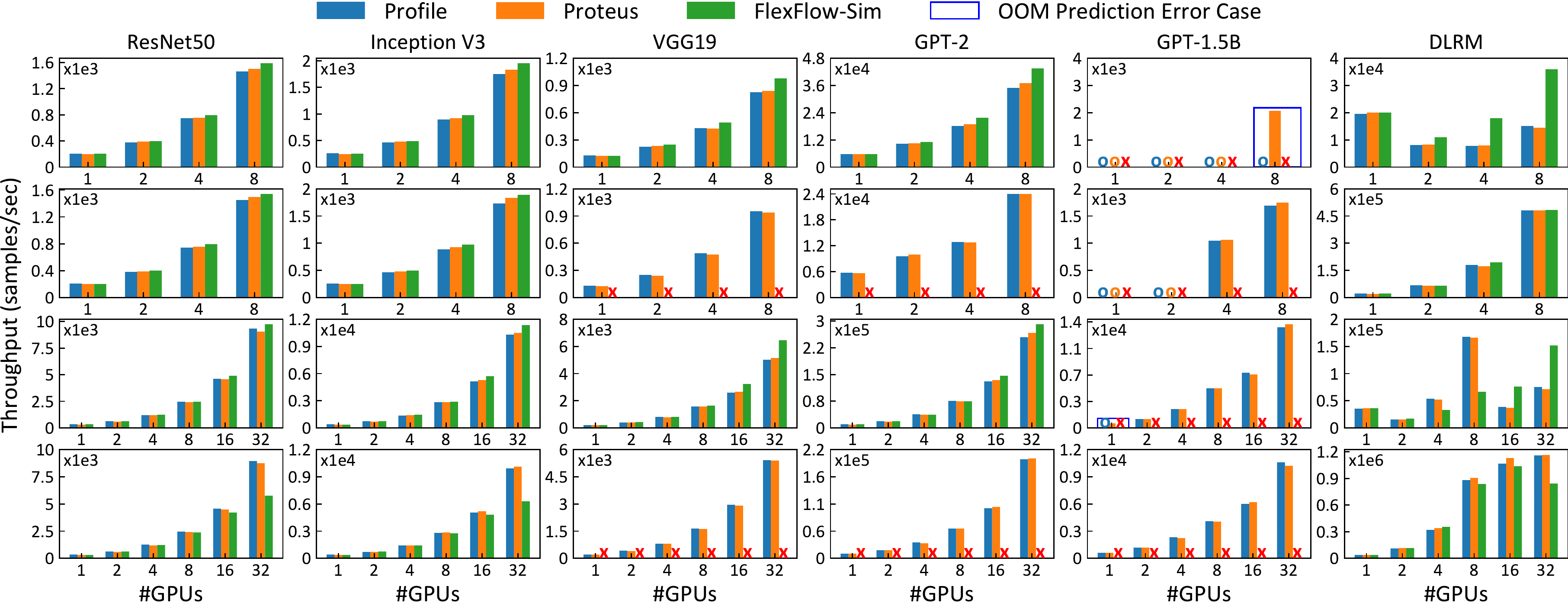}
    \caption{Throughput of DNN models on different hardware configurations with distinct parallelization strategies. The parallelization strategies from top to bottom row are $S1$ ($HC1$), $S2$ ($HC1$), $S1$ ($HC2$) and $S2$ ($HC2$), respectively. \textbf{o}: out of memory, \textcolor{red}{\textbf{$\times$}}: unsupported strategy.}
    \label{fig:sim_acc}
\end{figure*}

To evaluate \Sys across a wide variety of parallelization strategies, we evaluate each model with 2 popular parallelization strategies. One is most commonly used parallelization strategy ($S1$), the other is the optimal expert-designed parallelization strategy ($S2$). Since \Sys is aimed at accurate performance modeling rather than discovering new parallelization strategies, and implementing entirely new parallelization strategies is difficult, we do not evaluate \Sys on less commonly used parallelization strategies. But the parallelization strategies we tested already cover both operator- and subgraph-level strategies.

Since the most commonly used parallelization strategy is data parallelism, \Sys uses data parallelism or its variants as $S1$ for six DNNs. To enable data parallelism training of large model, \Sys combines memory optimization (ZeRO~\cite{ren2021zero}) and recomputation in $S1$ to evaluate GPT-1.5B. Expert-designed parallelization strategies ($S2$) exhibits more diverse patterns. ResNet50 and Inception\_V3 partitions data and output channels, while VGG19 and GPT-2 partitions data, output channels and reduction dimensions for computation parallelization. The $S2$ of GPT-1.5B combines op shard, pipeline and recomputation. DLRM partitions huge embedding table in $S2$ to optimize memory footprint.

\Cref{fig:sim_acc} shows the simulation results of various DNN models on two hardware configurations ($HC1$ and $HC2$) and \Cref{table:acc} displays the overall results on three hardware configurations. \Sys delivers an accurate performance model and achieves $3.0\%$ average prediction error for training throughput. Out of 180 simulation results, \Sys's estimation of OOM is incorrect only under 2 cases (blue box in \Cref{fig:sim_acc}).

\Sys is the first standalone simulator that targets on simulating complex parallelization strategies. The most related and representative cost-model and simulator is Paleo~\cite{qi2016paleo} and FlexFlow~\cite{jia2019beyond}, respectively. Paleo~\cite{qi2016paleo} is an analytical cost-model. It delivers high prediction error on single GPU (ResNet50 ($59.8\%$), Inception\_V3 ($40\%$)) and does not support GPT, DLRM models and complex parallelization strategies. Therefore, we did not dive into Paleo and show the results. FlexFlow~\cite{jia2019beyond} is an automated parallelization framework on SOAP space. To compare generated parallelization strategies, it internally tailors a simulator to simulate the training throughput. 

To compare \Sys and FlexFlow, we re-implement its simulator as FlexFlow-Sim. To support realistic simulation,  FlexFlow-Sim inserts collective communication operators for strategy transformation instead of point-to-point operators as described in FlexFlow paper. The comparison results are shown in \Cref{fig:sim_acc} and \Cref{table:acc}. The average prediction error for FlexFlow is $12.4\%$, which is $9.4\%$ higher than \Sys. Among all the test cases, the maximum error is $14.7\%$ and $137.9\%$ for \Sys and FlexFlow, respectively. For the total 180 training tasks, FlexFlow fails to estimate the performance of 1/3 of them. \Cref{fig:sim_acc} also shows that the prediction error of FlexFlow-Sim becomes larger as the number of GPUs increases.

We find that \Sys outperforms FlexFlow mainly in three aspects. 1) \Sys can be applied to a much larger parallelization strategy space with the abstraction of strategy tree. 2) FlexFlow ignores complex runtime behaviors thus cannot accurately model the training throughput. 3) FlexFlow's communication bandwidth estimation ignores fine-grained cluster topology. For example, FlexFlow delivers high prediction error for DLRM model, where communication dominates.

\begin{table}[t]
\centering
\small
\caption{Comparison of average and maximum prediction error of \Sys and FlexFlow-Sim (FF-Sim). Each strategy contains 15 results on 3 hardware configurations.}
\label{table:acc}
\scalebox{0.93}{
\begin{tabular}{ lccccc }
\hline
\multirow{2}{*}{Model} & \multirow{2}{*}{Strategy} & \multicolumn{2}{c}{Avg Error (\%)} & \multicolumn{2}{c}{Max Error (\%)} \\
& & \Sys & FF-Sim & \Sys & FF-Sim \\
\hline
\multirow{2}{*}{ResNet50} & \textit{S1} & 2.09 & 3.59 & 6.00 & 8.69 \\
 & \textit{S2} & 2.30 & 5.98 & 5.77 & 35.65 \\
\hline
\multirow{2}{*}{Inception\_V3} & \textit{S1} & 3.24 & 5.53 & 7.52 & 11.71 \\
 & \textit{S2} & 3.19 & 6.57 & 7.97 & 36.73 \\
\hline
\multirow{2}{*}{VGG19} & \textit{S1} & 1.97 & 8.11 & 4.97 & 28.17 \\
 & \textit{S2} & 1.68 & \ding{55} & 6.64 & \ding{55} \\
\hline
\multirow{2}{*}{GPT-2} & \textit{S1} & 2.56 & 6.97 & 6.20 & 24.14 \\
 & \textit{S2} & 2.31 & \ding{55} & 11.38 & \ding{55} \\
\hline
\multirow{2}{*}{GPT-1.5B} & \textit{S1} & 3.91 & \ding{55} & 8.09 & \ding{55} \\
 & \textit{S2} & 3.65 & \ding{55} & 9.92 & \ding{55} \\
\hline
\multirow{2}{*}{DLRM} & \textit{S1} & 5.07 & 48.14 & 14.68 & 137.89 \\
 & \textit{S2} & 4.55 & 14.05 & 11.44 & 114.63 \\
\hline
\end{tabular}
}
\end{table}

\subsection{Parallelization Strategy Comparison}

Comparing the training throughput of various parallelization strategies is an important problem in designing and understanding high performance parallelization strategies. In this section, we use GPT-2 as benchmark because GPT model is the most popular and widely used model to study all kinds of parallelization strategies and these strategies can generalize to other models. In these experiments, we select 4 parallelizable dimensions across operator- and subgraph level strategies and represent the parallelization strategy as $DP\times MP \times PP (n\_micro\_batch)$ and $DP$, $MP$ and $PP$ is the degree of data, model and pipeline parallelism. The global batch size is 8 and 64 for \textit{HC1} and \textit{HC2} respectively.

\begin{table}[t]
\centering
\small
\caption{The prediction error and rank of throughput for GPT-2 with different parallelization strategies on \textit{HC1} and \textit{HC2}. Rank is denoted in the format of \textit{truth / predicted} rank.
}
\label{table:trc_hc}
\begin{tabular}{ ccc|ccc }
\hline
\rowcolor{lightgray}
\multicolumn{3}{c|}{\textit{HC1}} & \multicolumn{3}{c}{\textit{HC2}} \\
Strategy & Error & Rank & Strategy & Error & Rank \\
\hline
8$\times$1$\times$1 ($1$)  & $2.34\%$  & 2 / 2 & 16$\times$1$\times$1 ($1$) & $3.38\%$ & 1 / 1 \\
4$\times$2$\times$1 ($1$)  & $2.92\%$  & 1 / 1 & 8$\times$2$\times$1 ($1$)  & $2.89\%$ & 2 / 2 \\
2$\times$4$\times$1 ($1$)  & $2.43\%$  & 3 / 3 & 4$\times$4$\times$1 ($1$)  & $5.87\%$ & 3 / 3 \\
1$\times$8$\times$1 ($1$)  & $1.77\%$  & 5 / 5 & 2$\times$8$\times$1 ($1$)  & $3.88\%$ & 4 / 4 \\
2$\times$2$\times$2 ($1$)  & $2.54\%$  & 6 / 6 & 8$\times$1$\times$2 ($4$)  & $3.60\%$ & 6 / 6 \\
2$\times$2$\times$2 ($2$)  & $1.60\%$  & 4 / 4 & 8$\times$1$\times$2 ($8$)  & $2.99\%$ & 5 / 5 \\
                           &           &       & 2$\times$4$\times$2 ($4$)  & $5.64\%$ & 7 / 7 \\
\hline
\end{tabular}
\end{table}

Table~\ref{table:trc_hc} shows the simulation results of GPT-2 with various parallelization strategies. For these parallelization strategies, \Sys can accurately model the performance and achieves $3.2\%$ average prediction error. Order preservation is an important feature in strategy comparison and \Sys maintains the rank of diverse parallelization strategies. \Cref{table:trc_hc} demonstrates that \textit{HC2} prefers data parallelism training, since hybrid model parallelism shares the bandwidth of IB net, and pipeline parallelism introduces bubbles during training, which will decrease the training throughput. The simulation results also confirms that pipeline efficiency can be improved by injecting more micro batches. \textit{HC1} consists of a single NUMA node, and the 2-way model parallelism can fully utilize the QPI links between two CPU sockets, thus achieving highest throughput.

\begin{figure}[t]
    \centering
    \includegraphics[width=0.48\textwidth]{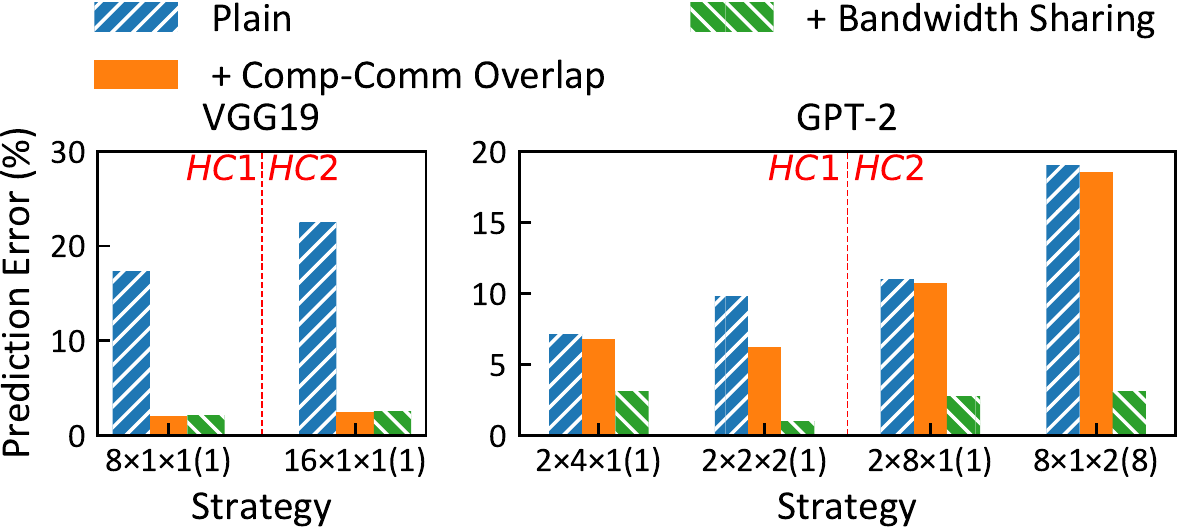}
    \caption{The prediction error of VGG19 and GPT-2 with different components. Plain: \Sys without runtime behavior detector.}
    \label{fig:ablation}
\end{figure}

\subsection{Runtime Behavior Ablation Study}
\label{ablation}

To study the effective of \textit{runtime behavior detector}, we test the throughput of VGG19 and GPT-2 on \textit{HC1} and \textit{HC2} with different parallelization strategies. For VGG19, we use batch size 32 per GPU with data parallelism training. For GPT-2, the global batch size is 8 and 64 on \textit{HC1} and \textit{HC2} with hybrid op shard and pipeline parallelism. \Cref{fig:ablation} shows that \textit{runtime behavior detector} can greatly improve the simulation accuracy of throughput (average error: Plain ($14.4\%$) vs Proteus ($2.4\%$)). VGG19 is very sensitive to \textit{comm-comp overlap}, hence introducing overlap factor can significantly improve the prediction accuracy. As there is no \textit{bandwidth sharing} in data parallelism training, prediction error of VGG19 keeps after adding \textit{bandwidth sharing}. In contrast, GPT-2 is more sensitive to \textit{bandwidth sharing} which is especially common in complex parallelization strategies. Therefore, the throughput prediction error reduces remarkably after modeling \textit{bandwidth sharing}.

\subsection{Simulation Cost}

To evaluate the simulation cost of \Sys, we measure the time it takes to evaluate VGG19 and GPT-2 on \textit{HC2} with data parallelism. Since the cost of computation operators can be profiled in advance, we only evaluate the time cost of \textit{execution graph compiler} and \textit{HTAE}. \Cref{table:cost} demonstrates that \Sys takes seconds to simulate the performance of DNNs with a large number of GPUs. We believe this cost is acceptable to evaluate a specified parallelization strategy since \Sys provides a fine-grained simulation without requiring GPU resources. In contrast, profiling a general parallelization strategy will take a lot of effort and GPU resources.

\begin{table}[t]
\centering
\small
\caption{Simulation cost of \Sys on \textit{HC2} in \textit{seconds}.}
\label{table:cost}
\begin{tabular}{ c|\cX{0.9cm}\cX{0.7cm}\cX{0.7cm}|\cX{0.9cm}\cX{0.7cm}\cX{0.7cm} }
\hline
\multirow{2}{*}{\#GPUs} & \multicolumn{3}{c|}{VGG19} & \multicolumn{3}{c}{GPT-2} \\
\cline{2-7}
 & compile & exe. & total & compile & exe. & total \\
\hline
1  & 0.033	& 0.006 &  0.039 & 0.188	& 0.070 & 0.258 \\
2  & 0.053	& 0.407 &  0.460 & 0.341	& 0.450 & 0.792 \\
4  & 0.114	& 0.530 &  0.644 & 0.504	& 0.692 & 1.196 \\
8  & 0.170	& 0.563 &  0.733 & 1.008	& 0.873 & 1.881 \\
16 & 0.336	& 0.630 &  0.966 & 1.966	& 1.172 & 3.138 \\
32 & 0.777	& 0.921 &  1.698 & 4.143	& 2.123 & 6.265 \\
\hline
\end{tabular}
\end{table}
\section{Related Work}

\noindent
\textbf{Handcrafted Parallelization Strategies} are designed to optimize distributed DNN training.
One wired trick~\cite{krizhevsky2014one} introduces model parallelism for linear layers to accelerate AlexNet. Megatron-LM~\cite{shoeybi2019megatron} presents an expert-designed strategy to expedite transformer models combining data, model and pipeline parallelism. DeepSpeed~\cite{rasley2020deepspeed} introduces ZeRO to reduce memory footprint by partitioning model states across data parallel processes. Recomputation \cite{chen2016training} utilizes tensor rematerialization to decrease memory consumption. \Sys~can model the performance of these manual designed strategies thus assisting their analysis and optimization.

\noindent
\textbf{Automatic Parallelization.}
FlexFlow~\cite{jia2019beyond} and Tofu~\cite{wang2019supporting} proposes SOAP and partition-n-reduce space to parallelize operators. GSPMD~\cite{xu2021gspmd} introduces a more general parallelization space by partitioning all parallelizable dimensions of tensors.
DAPPLE~\cite{fan2021dapple} and PipeDream~\cite{narayanan2019pipedream} optimize parallelization strategies in data and pipeline parallelization space. Alpa~\cite{zheng2022alpa}
combines data, model and pipeline parallelism and proposes a inter-operator and intra-operator
parallelization space. Existing automatic approaches focus on exploring the space of computation parallelization, while our work introduces a unified parallelization strategy space considering computation parallelization and memory optimization at operator level and schedule at subgraph level.

\noindent
\textbf{Performance Model.}
Previous works propose analytical performance models for DNN training on single GPU  \cite{liu2021seer, kothapalli2009performance, zhang2011quantitative}
or on multiple GPUs with data parallelism or hybrid data and model parallelism \cite{yan2015performance, qi2016paleo, pei2019iteration}. These approaches are not applicable to increasingly complex training workload and strategies. FlexFlow~\cite{jia2019beyond} introduces a simulation model to estimate the cost of a SOAP strategy, but it is not designed to capture the cost of general strategies and runtime behaviors. \Sys aims to provide a general simulation performance model for various parallelization strategies.

\section{Conclusion}
In this work, we present \Sys to simulate the performance of distributed DNN training strategies on diverse clusters. \Sys~features \textit{strategy tree} to model unified parallelization strategy space and \textit{hierarchical topo-aware executor} to model runtime behaviors of computation and communication operators accurately. We can leverage \Sys to analyze and optimize the performance of general parallelization strategies.



\bibliographystyle{IEEEtran}
\normalem
\bibliography{bib/ref.bib}

\vfill

\end{document}